# Modeling trading games in a stochastic non-life insurance market

Leonard Mushunje[1*], David Edmund Allen[2]

[1]Columbia University, Department of Statistics

[2]University of Sydney, School of Mathematics

lm3748@columbia.edu

**Abstract**

We studied the behavior and variation of utility between the two conflicting players in a closed Nash-equilibrium loop. Our modeling approach also captured the nexus between optimal premium strategizing and firm performance using the Lotka-Volterra completion model. Our model robustly modeled the two main cases, insurer-insurer and insurer-policyholder, which we accompanied by numerical examples of premium movements and their relationship to the market equilibrium point. We found that insurers with high claim exposures tend to set high premiums. The other competitors either set a competitive premium or adopt the fixed premium charge to remain in the game; otherwise, they will operate below the optimal point. We also noted an inverse link between trading premiums and claims in general insurance games due to self-interest and utility indifferences. We concluded that while an insurer aims to charge high premiums to enjoy more, policyholders are willing to avoid these charges by paying less.

## 1. Background and Introduction

Technological advances and other related influential driving factors have made and are still making our economies unstable and ever-changing environments. We find some economic volatility traits troublesome to various economic agents. From another angle, most investors, if not all of them, are risk averse, which is interesting in terms of business successes. While others are risk averse, some favorably identify business opportunities that find risks as viable business prospects. All this backs up the present-day existence of the insurance and risk management industry. Insurance is increasingly becoming a particular area of interest in the finance industry.

Generally, insurance is the pooling of risks. It is a deal between two independent parties whose interests and motives will differ but with a sense of dependability, CFPB, (2022). Here, one part is known to contribute monthly payments known as premiums, and the other will be responsible and answerable for the paid premiums in case of a loss or risk. So, risks are transferred from one part to the other, which we refer to as risk pooling and, shortly, insurance. Therefore, insurance is about the interaction between the insurers (insurance firms) and the policyholders (clients). These interactions are of beneficial interest, and one needs to understand them well, forming the basis of our paper. Insurance is associated with economic cycles, and insurance players often find opportunities to maximize their payoffs at the expense of their policyholders. At the same time, during cycles, policyholders tend to claim more at the expense of the insurers' returns, Wakif A. et al., (1981); Lamm-Tennant et al., (1997); Enz, R. (2002).

In this paper, we aim to extract, examine, and model the possible interacting behavior of insurers and their policyholders 'counterparts within the insurance industry regarding their risks and returns. Insurance is a gaming system; this paper is motivated to theoretically model business games in the insurance industry. Interestingly, we considered the insurance industry as an ecosystem where various interactions occur for the actors' main survival, taking insures and the policyholders as the main actors of the case. Shi and Sheng (2008) in their note said, "When we observe our little garden, we could see spiders, ambush bugs and pray mantis to hide silently for their victims, while other little butchers, such as dragonflies and tiger beetles, actively move around to find their food sources." The same applies to the insurance ecosystem; it is possible to find different clients and many insurers of different aspects and sections of businesses, and their behavioral ways and interactions are of noble interest. Therefore, this paper aims to examine the games played in the insurance industry, where we shall look at how risks and returns move and behave in the insurance sector. Since we shall define our actors regarding risks and returns, we must define them all.

Risks are adverse conditions that mostly bring forth unfavorable operating conditions and failures, whereas returns are payoffs that typically follow any investment action. It is a commonly known investment phenomenon where risks and returns may appear, forming the baseline for our paper. While different Game theory models have been used, especially in economics, this paper used another different approach. We shall use the Predator-Prey model built from the differential equation idea and some basic assumptions. For interest and viability sake and to reach our consensus, we shall also use a Prey-Predator model. Therefore, we shall use a two-way approach that declares a sense of fairness. The model is widely used in the biological field, and literature on the subject exists. However, its application to the insurance sector must be revised, reinforcing our study's novelty.

The Game theory model should be appropriate for this problem because predation is played between predator and prey. However, more research needs to be done on how predation and anti-predation behavior evolve for predators and prey with different body sizes. Despite insufficient research, other research icons put some sounding-related materials on paper. Chen, Xu, and Guo (2018) discussed a predator-prey system with distributed delay in biology. They proposed a discrete model that enabled them to derive a globally asymptotic stable positive periodic solution for the studied organisms. Zhang and Hou (2010) investigated a ratio-dependent predator-prey system's four positive periodic solutions, almost similar to Chen et al. We also have other writers, such as Liu (2014) and Dunkel (1968), who implemented oscillatory dynamics of predator-prey systems in the biological fields. Other exciting work by significant authors, such as the one by Chakraborty, Pal, and Bairagi (2011), exists. The team studied a simple predator-prey model where the predator population is subjected to harvesting. They used such a model mathematically with biological ramifications, and the results included switching off stability, oscillations, and deterministic extinction. In their note, they pointed out a helpful point: In theoretical ecology, two primary functional responses are prey-predator and predator-dependence. The most recent work on the application of the Lotka Volterra model in ecology can be found in Pal et al. (2019a), Pal et al. (2019b), Basaznew B, and Dawit M (2022).

Our core aim is to apply the same functions in fields like insurance and other interesting, non-ecological ones. Mathematical research on the predator-prey model was done widely, except on the ecological subject. Writers such as Kuang and Beretta (1998), Tang and Zhang (2005), and Xiao and Ruan (2001) have related exciting works on the model that is forming the basis of our interest. Game theory in ecology has been widely researched, and the list of references is almost endless. However, we must acknowledge the works of Xiao, Li, and Han (2006) to write our literature. The three systematically studied the dynamical properties of a ratio-dependent predator-prey model with non-zero constant rate predator harvesting. Their results revealed that the model has two equilibria in the first quadrant and can exhibit numerous phenomena, including the bifurcation of the cusp type of co-dimension.

**Literature on Game Theory in Insurance Business**

Game theory application in insurance dates back as marked by several authors' and practitioners' work on the subject. We see extensive work on applied cooperative games and insurance risk conveyance by Borch, 1962, 1974), Buhlmann, 1980, 1984), and Lemaire, 1984, 1991). This is related to the broad extended work by (Aase, 1993; Brockett & Xia, 1995; Tsanakas & Christofides, 2006). The Bertrand oligopoly and Cournot oligopoly models were applied in the non-life insurance business through non-corporative Game theory, where insurers can opt for the business volumes they want. Relatedly, Boonen (2016) put forward an approach to control reinsurance risks optimally. Under a single-time model, the author looks at the interaction of different premium strategies of two insurers in the trade of risks in the counter OTC) markets. Related to our work is the contribution from Emms (2012), who applied differential games to non-corporative insurance markets. The underlying assumption is that insurers are trading at optimal strategic points/le; other insurers' premiums depend on their competitor's thresholds.

On the other hand, Wu and Pantelous (2017) studied the notion of game aggregation using the one-time horizon approach. The framework aggregates all players induced through payoff pairing. Other exciting work is seen in (Monderer and Shapley, 1996; Jensen, 2010). Other scholars such as Friedman (1987), Ruan (2009), and Bohner, Fan, and Zhang (2006) exist in literature with the same message of games and competition among units. To the best of our knowledge, our model has never been applied to the literature, but our analysis methods share some traits with that of Chakraborty, Pal, and Baragi (2012). The novelty of their study was that they studied the system concerning two controllable parameters and discussed the results from the ecological point of view. This, combined with other works, also created our research gap and the novelty of our study.

Our paper shall apply a non-ratio two-way competition model (predator-prey and prey-predator model) in the insurance sector to analyze and discuss the theoretical games between insurers and policyholders regarding their respective returns and risks of trading. Our objective is to derive an internal stable coexistence point where both parties can operate within the insurance market and suggest possible ways of avoiding chances of extinction. Therefore, several theoretical mathematics shall be applied and conducted with robust mathematical analysis to reach our desirable conclusions.

## 2. Methodology

### 2.1. Models

As we are concerned with investigating the interactions between insurers and policyholders in the insurance field, we tend to base our focus firstly on games based on premium setting (pricing) and secondly on risk and return trade-off in terms of the policyholders and insurers, respectively, to see how policyholder's risks behave in response to their insurers' counterparts. All these scenarios are captured by different but dependent models defined subsequently. Since the study aims to capture the competition between two or more players, we opted to apply the Lotka Volterra model, a widely known model for ecosystem competition modeling.

### 2.2. Finite-Continuous time Lotka-Volterra competition and cooperative differential game models: A new approach

We formulated finite-time Lotka-Volterra competition and cooperative differential games with a finite number of players (insurers and policyholders). From the model, we assessed the strength of each player's strategy against the derived NASH Equilibrium points, which we call differentiated NASH Equilibrium points. Ideally, altering any premium when playing/trading at these NASH points is impossible. Our basis inference is that players implement optimal strategies at equilibrium. There are no other immediate optimal strategies since we are assuming an imperfect market with finite players. This leads to market stabilization, mainly in the long run.

Our model is a competitive typed model where, in this case, insurers are competing for a shared resource. Insurers are competing for high expected premiums and high returns. On the other hand, policyholders are competing for low prices/premiums from insurers. Thus, the model captured/accommodated market players from the same group and different groups (insurers-insurers, Insurers-Policyholders, Policyholders-Policyholders). The invention of a cooperative model captures the games where the players agree and cooperate in the market. The Lotka-

Volterra model is based on the exponential changes in population, but in this case, we considered a competitive model of a logistic nature. Our analysis considered two-player games consisting of two insurers, an insurer, and a policyholder. For simulations, we generalized our model up to n-number of players, where in our case, n is 10.

**2.2.2. Two-player competitive game model**

In this case, our model consists of four variables: $N(t), K, \rho,$ and $r$, representing the total number of players at time t, market threshold, growth rate, and the intraspecific competition over economical premiums among the players, respectively. The resulting model becomes:

$$\frac{dN}{dt} = \rho N \left(1 - \frac{N}{K}\right), K > 0, [1].$$

From the model defined above, we measure the aggressiveness of the competition in the trading game. We then write our new model with $c_1$ and $c_2$ as the competition intensity.

$$\frac{dN_1}{dt} = \rho_1 N_1 \left(1 - \frac{N_1}{K_1} - c_1 N_2\right), [2]$$

$$\frac{dN_2}{dt} = \rho_2 N_2 \left(1 - \frac{N_2}{K_2} - c_2 N_1\right), [3].$$

Measuring competition intensity among insurers at the edge to provide general insurance services is fundamental. Policymakers in Zimbabwe (IPEC) may be keen to control and set rules against all insurers based on the nature of their trading associations and behavior.

**2.2.3. Two-player Cooperative Model**

In economic terms, the players (insurers and policyholders) can cooperate in trading. Here, we also study a case of corporative games in the context of insurers' games on pricing and premium settings. We considered players of the same characteristics (same group), that is, insurers in the general insurers. The model employed in this section is analogous to the LV system described above, with minor differences only in the signs of the interaction terms. We provided the following definitions, which are equally helpful in this section. We adopt them from the work of Mafalda (2017).

Definition 1: A cooperative matrix is any real n × n matrix with a sign structure.

Definition 2: Negatively diagonally dominant: A matrix A is negatively diagonally dominant if $d \in R^n: d_i > 0$ exists and $a_{ii}d_i + \sum_{i \neq j}|a_{ij}d_j| < 0, \forall = 1, \ldots, n$. We have $(Ad)_i < 0, \forall i = 1, \ldots, n$ for cooperative games.

Definition 3: Stable Matrix: an $n \times n$ matrix with a strictly negative real part on all its eigenvalues.

Now, having defined all the relevant vital elements, we present our model as follows:

$$\frac{dN_1}{dt} = \rho_1 N_1 \left(1 - \frac{N_1}{K_1} + c_1 N_2\right), [4]$$

$$\frac{dN_2}{dt} = \rho_2 N_2 \left(1 - \frac{N_2}{K_2} + c_2 N_1\right), [5]$$

Like in our competitive model, we simplified our model as follows:

$$\frac{du_1}{dT} = u_1(1 - u_1 + a_{12}u_2), [6]$$

$$\frac{du_2}{dT} = \rho u_2(1 - u_2 + a_{21}u_1), [7]$$

with, $u_1 = \frac{N_1}{K_1}; u_2 = \frac{N_2}{K_2}; a_{12} = c_1 K_2; a_{21} = c_2 K_1; T = \rho_1 t; \rho = \rho_1 \rho_2$.

### 2.2.4. Insurers-Policyholders' Game Model: Inverse transactions

Here, we analyze games played by insurers and policyholders themselves. The games bring utility theory in that each player aims to maximize utility from the games/trades done over the counter and through regulated platforms. Hence, in this section, games regarding risks and returns corresponding to policyholders and insurers are analyzed, respectively.

Model Assumptions.

We used some assumptions during the modeling framework, including:

Let P(t) and R(t) be the policyholder's risks and insurers' returns at time t, respectively; Stable free economic conditions; Perfect information that is zero market failures. The prey (policyholders) is divided into two groups: the active and passive policyholders. There are no other sources of income for insurers except premiums; There are no other risk transfer pots for policyholders except insurers. Both insurers and policyholders are the market controllers (leaders) depending on the game advantage: Permeable trading zones. The-model is uniquely structured in that predators represent insurers while the prey is the policyholders, but we considered both the insurer's returns and the policyholder's risks under restricted conditions.

Zero-interaction condition

$$\frac{dP}{dt} = \delta p, [8]$$

$$\frac{dR}{dt} = -\alpha r, [9], \text{ where}$$

P is the policyholder's risk quantity, $\delta$ is the risk growth rate, R is the insurer's return quantity at time t, and $\alpha$ is the return decay rate.

The above two equations hold and are well defined under some conditions, such as $\delta > 0$, $\alpha > 0$, and $t \geq 0$. The gaming system in the insurance sector is identified by the view and appreciation of the growth behavior and interrelations of these two competing and inverse components—non-zero interaction conditions.

An ecosystem consists of interdependent organisms of different types. Likewise, an insurance ecosystem consists of insurers, policyholders, brokers, actuaries, underwriters, and regulators. In this study, insurers and policyholders are considered the key players in the insurance sector, driven by their interdependence traits. Therefore, for sufficiency and gaming fundamentals, the special second case is defined below,

$$\frac{dP}{dt} = \delta p - \varepsilon pr, [10]$$

$$\frac{dR}{dt} = \alpha pr - \beta r, [11] \text{Where,}$$

$\delta$ is the policyholder's risk growth, $\varepsilon$ is the interdependence coefficient between risks and returns for policyholders and insurers, $\beta$ is the insurer's return decay rate, and $\alpha$ is the interdependence coefficient between risks and returns for policyholders and insurers.

3. **Results and findings**

3.1. **Competitive model results**

To analyze the solution of our model, we have two possible cases. Firstly, we take either $N_1(0)$ or $N_2(0)$ to be zero. This gives us a single insurer/player model, and with time, that particular insurer goes to either extinction or carrying capacity at an exponential rate:

$$N_{10} = 0 \implies N_1(t) = 0, N_2(t) = \frac{N_{20}}{\frac{N_{20}}{K} + \left(1 - \frac{N_{20}}{K}\right)e^{-\rho_2 t}}$$

$$N_{20} = 0 \implies N_2(t) = 0, 1(t) = \frac{N_{10}}{\frac{N_{10}}{K} + \left(1 - \frac{N_{10}}{K}\right)e^{-\rho_1 t}}$$

After solving the model, we get the following Jacobean function:

$$J = \begin{bmatrix} 1 - 2u_1 - a_{12}u_2 & -a_{12}u_1 \\ -\rho a_{21} & \rho(1 - 2u_2 - a_{21}u_1) \end{bmatrix},$$ [12] which aids in deriving the equilibrium points, which we call New NASH equilibrium points that is, $(0, 0); (1, 0); (0, 1);$

$P^* = \left(\frac{1 - a_{12}}{1 - a_{12}a_{21}}, \frac{1 - a_{21}}{1 - a_{12}a_{21}}\right)$. Secondly, we assure the non-negativity of the number of players from the sets of steady points obtained among the NASH points. Below are the cases involving such analysis, where stability analysis of each point is considered consciously. Stability analysis greatly depends on the choice of parameters, as we shall note immediately.

i). A: $a_{12}, a_{21} < 1$; While (0, 0) is not stable, the other points (1, 0) and (0, 1) are saddles, but they ensure the non-negativity of the populations. However, our steady point at P is stable. Small parameter values ensure less aggression and minimal competition among the insurers; hence, we observe an equilibrium coexistence between the players/insurers in competition. See Figure 3.5.

ii). B: $a_{12}, a_{21} > 1$; Unlike in case 1, our steady point and (0, 0) are unstable, with both (1, 0) and (0, 1) stable, and the resulting plane is divided into two sections by an almost faint line. We have trajectories approaching (1,0) above this boundary line, with the ones below approaching (0,1). Some orbits that originate from the line approach the unstable steady point at P. Ideally, increasing the magnitude of the parameter values results in massive interspecific competition among the players. As shown in Figure 3.6, the winner of the game is determined by the comparative starting advantage iii). C: $a_{12} < 1, a_{21} > 1$; we have no steady point at P. Instead, we have nodes (0,0) and (0,1) as unstable, with only (1,0) as the stable state. All our trajectories move toward these state points, as in Figure 3.7. The parameter value $a_{12}$ gives information on the competition intensity among players.

iv). D: $a_{12} > 1, a_{21} < 1$; Like in step three above, we do not have an interior steady state. It is precisely the opposite for the rest of the points. The nodes (0, 0) and (1, 0) are unstable, and (0, 1) are stable. All trajectories move toward the stable state (0,1); see Figure 3.8.

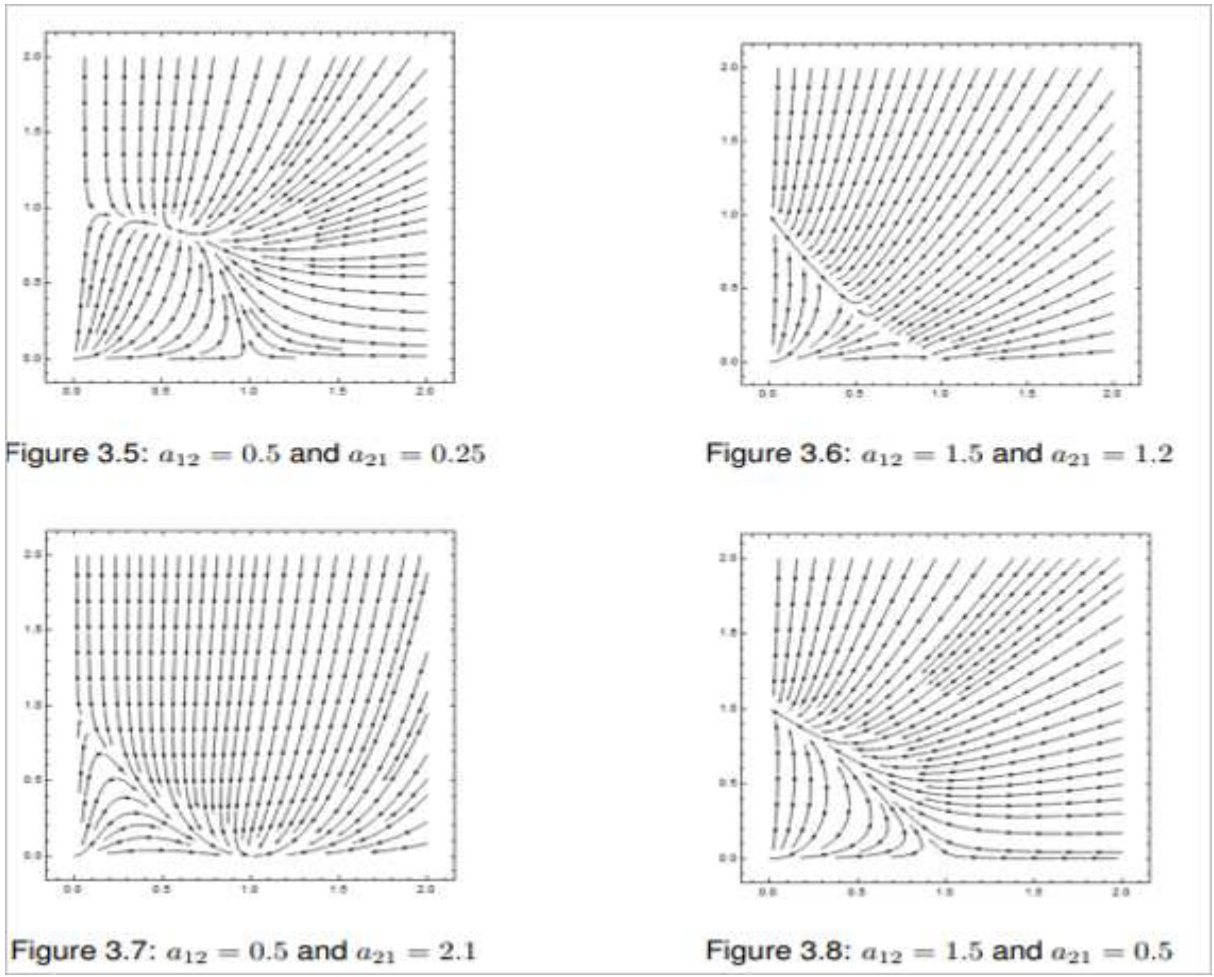

Figure 3.5: $a_{12} = 0.5$ and $a_{21} = 0.25$

Figure 3.6: $a_{12} = 1.5$ and $a_{21} = 1.2$

Figure 3.7: $a_{12} = 0.5$ and $a_{21} = 2.1$

Figure 3.8: $a_{12} = 1.5$ and $a_{21} = 0.5$

We note that players with corresponding small parameter values are more competitive and have the chance to control the market. They can enjoy their competitive advantage, as they can, to a greater extent, set the premiums others can follow.

### 3.2. Cooperative model results

Like in competitive games, the steady point, P, occurs when $a_{12}, < 1$, ensuring the favorable population of the insurers. From the model analysis, we have the following signed Jacobean matrix. So, cooperative games have a signed J-matrix false for their competitive counterparts.

$$J = \begin{bmatrix} 1 - 2u_1 + a_{12}u_2 & a_{12}u_1 \\ \rho a_{21} & \rho(1 - 2u_2 + a_{21}u_1) \end{bmatrix}, [13]$$ which aids in the stability analysis of the following equilibrium points:

$(0, 0); (1, 0); (0, 1);$

$$P^* = (\frac{1+a_{12}}{1-a_{12}a_{21}}, \frac{1+a_{21}}{1-a_{12}a_{21}}).$$

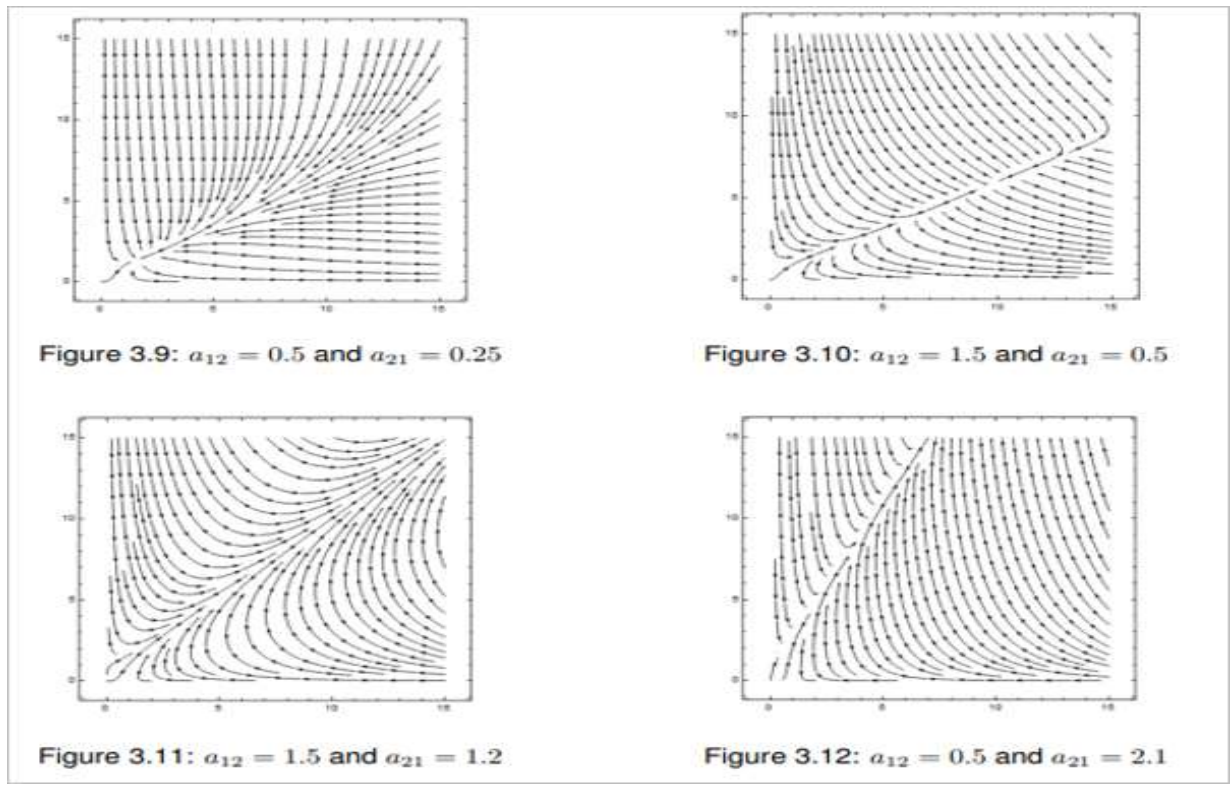

Figure 3.9: $a_{12} = 0.5$ and $a_{21} = 0.25$

Figure 3.10: $a_{12} = 1.5$ and $a_{21} = 0.5$

Figure 3.11: $a_{12} = 1.5$ and $a_{21} = 1.2$

Figure 3.12: $a_{12} = 0.5$ and $a_{21} = 2.1$

Using our Jacobean matrix, we analyzed the cooperative NASH equilibrium points in three cases.

i). A: $a_{12}a_{21} < 1$; The small parameter values guarantee the existence of a steady point, P, from which the trajectories converge. See Figures 3.9 and 3.10.

ii). B: $a_{12}a_{21} > 1$; We have (0, 0), (0, 1), and (1, 0) as our nodes, with no interior steady point. The absence of this node results in infinity divergence of the orbits and corresponding trajectories, as in Figures 3.11 and 3.12. This is true for any values of the parameters in the defined domain.

### 3.3. Inverse transaction model results

Considering the predator-prey model above, we get the following

$P = Ae^{\delta t}$ ∗∗and $R = Be^{-\alpha t}$ ∗∗∗ represents the solution for zero interaction between insurers and policyholders.

Our interpretation for both solutions is that A is the upper asymptote for the policyholders' zero interaction solution, and $\delta$ is the risk growth factor defining the shape parameter. The same applies to B and $\alpha$ for the return solution, except that $\alpha$ is a decaying factor. Both solutions ** and *** represent the growth behavior of risks and returns for policyholders and insurers, respectively. The results show no games between insurers and policyholders without interaction. From such, the policyholder's risks grow logistically without any sense of extinction. Even if the risks and returns decrease or increase, zero values for the two cannot be obtained. We can safely find the maximum risks a policyholder can realize and the minimum possible return that insurers can realize. This is found by dividing the upper/lower limit with the logarithmic base, e, as illustrated below. $K_{max/min} = \frac{A}{e}$ is helpful as it is used as a guideline in decision-making by main actors like, in this case, insurers and policyholders. Their gaming actions do follow their exceptional knowledge of this particular constant value. Interestingly, the same formula is applied to the non-zero interaction conditional gaming case. We shall now look at the non-zero interaction case.

Non-zero interaction problem and Equilibrium and stability analysis

We found four distinct sets of critical points and further examined their stability/instability nature using the J matrix.

I. Predator-prey free state $(0,0)$.
II. Predator-free state $(p^*, 0)$
III. Prey free state $(0, r^*)$
IV. The coexistence state $(p^*, r^*)$.

More specifically, we have the actual points as follows:

I. Equilibrium free state $(0,0)$.
II. Predator free point $\left(\frac{\delta}{\alpha}, 0\right)$
III. Prey free point $\left(0, \frac{\delta}{\varepsilon}\right)$.
IV. Coexistence point $\left(\frac{\beta}{\alpha}, \frac{\delta}{\varepsilon}\right)$.

The nature of the above points is determined to thoroughly examine the relational links between risks and returns for policyholders and insures during their games in the insurance field. The following is a stability analysis obtained from this system:

We performed our stability analysis separately for the two sets of four equilibrium points obtained from our models, starting with the predator-prey model. Like before, we used the Jacobean approach, which uses a Jacobean matrix to find the eigenvalues from which the stability/instability of each point is deduced. In some cases, this is accompanied by a system of fitted trajectories. Before our analysis, we define a set of conditions we will use. Thus, $0 < (\delta, \alpha, \varepsilon, \beta) \leq 1$ and $\delta > \beta$, $\alpha > \varepsilon$. Now, the equilibrium points from the predator-prey model are:

$(0,0)$, $\left(\frac{\delta}{\alpha}, 0\right)$, $\left(0, \frac{\delta}{\varepsilon}\right)$ and $\left(\frac{\beta}{\alpha}, \frac{\delta}{\varepsilon}\right)$, and we singly analyze them as:

The Jacobean matrix is given by:

$$J = \begin{pmatrix} f_p & f_r \\ g_p & g_r \end{pmatrix}, [14]$$ and the eigenvalues are given by,

$$|J - \lambda I| = 0$$

So, we let $f = \delta p - \varepsilon p r$ and $g = \alpha p r - \beta r$ such that,

$f_p = \delta - \varepsilon r$

$f_r - \varepsilon p$

$g_p = \alpha r$

$g_r = \alpha p - \beta$. Now for $(0,0)$ we get:

$$\begin{vmatrix} \delta - \lambda & 0 \\ 0 & -\beta - \lambda \end{vmatrix} = 0 \text{ thus,}$$

$\lambda_1 = \delta$ or $\lambda_2 = -\beta$. These are two distinct and real values that characterize a saddle point. This is always an unstable point. This means that even in the long run, gaming in the insurance sector has zero chance of getting extinct point. There are no chances for insurers and insureds to get no returns and risk values known as co-integration operational level. Relating to the above trajectory nature, the trajectories move away from the point $(0,0)$.

Secondly, for $\left(\frac{\delta}{\alpha}, 0\right)$, after a procedural takeup, we finally get our reduced determinant function as:

$$\begin{vmatrix} \delta - \lambda & -\frac{\varepsilon \delta}{\alpha} \\ 0 & \delta - \beta - \lambda \end{vmatrix} = 0$$

$\lambda_1 = \delta$ and $\lambda_2 = \delta - \beta, \delta > \beta$ These are two positive eigenvalues for all values of δ and β, and they characterize a node. This unstable point implies that insurers and policyholders cannot play now. Thirdly, for $\left(0, \frac{\delta}{\varepsilon}\right)$ we will get the below finalized determinant matrix,

$$\begin{vmatrix} -\lambda & 0 \\ \frac{\alpha\delta}{\varepsilon} & -\beta - \lambda \end{vmatrix} = 0$$

$\lambda_1 = 0$ and $\lambda_2 = -\beta$. These are two distinct values that give only one eigenvector. This makes and classifies it as an improper and asymptotically stable point. However, finding existing insurers' returns with zero policyholders' risks in the insurance sector is an anomaly. This means that insurers cannot get returns from premiums without policyholders. However, this can be possible if both parties operate in the games' negative or slow-down field.

For $(\frac{\beta}{\alpha}, \frac{\delta}{\varepsilon})$, we get,

$$\begin{vmatrix} -\lambda & -\frac{\beta\varepsilon}{\alpha} \\ \frac{\alpha\delta}{\varepsilon} & -\lambda \end{vmatrix} = 0$$

This gives $\lambda^2 - \beta\delta = 0$, and finally, we have our eigenvalues as:

$\lambda_{1,2} = \pm\sqrt{\beta\delta}$. This coexistence point defines the normal and fair point of operation. However, in our case, the point is unstable, but it can be conditioned by making all the point values positive. Making them positive makes the point a stable operation node where both parties can operate profitably.

The obtained eigenvalues from all three other cases suggest a similar stability nature. The exact nature, however, depends on the gaming parties' values and region of operation. Nevertheless, the coexistence point is generally always stable, with the rest unstable. In a better mathematical and economical way, our derived points' exact and accurate nature depends on the region of operation within the gaming axils and the assigned corresponding values for the parameters used. We observed that the strength of each player's strategy could be explained through stability analysis of each of the derived NASH equivalent points. We noted that a player's strategy is considered superior to the opponent's if it is stable. The trajectories converge towards that critical point, and the opposite holds for weak strategies.

## 4. Numerical Simulations and Empirical Evidence

We provided some simulation results from our models to compute the risk premiums and risk exposures of each player involved in the game. We benchmarked our numbers using the non-life insurance market data from South Africa.

Table 4. 1: Nash-Equilibrium points: Premiums and Claim Exposures

**New NASH equilibrium premiums**

| Premiums | Insurer 1 | Insurer 2 | Insurer 3 | Insurer 4 | Insurer 5 | Insurer 6 | Insurer 7** | Insurer 8* | Insurer 9 | Insurer 10** |
|---|---|---|---|---|---|---|---|---|---|---|
| Nash Equilibrium (P**) | 197.45 | 191.87 | 203.11 | 208.75 | 199.03 | 186.78 | 167.89 | 216.14 | 195.34 | 181.93 |
| Indifference premium/Market premium | 180.45 | 176.88 | 195.13 | 200.94 | 183.22 | 175.33 | 175.57 | 183.33 | 186.77 | 191.98 |

**Corresponding Claim Exposures at NASH Equilibrium premiums**

| Claim Volumes | | | | | | | | | | |
|---|---|---|---|---|---|---|---|---|---|---|
| Nash Equilibrium (P**) | 6,802 | 7,542 | 8,302 | 7,431 | 7,087 | 6,473 | 1,001 | 9,796 | 4,827 | 1,319 |
| Indifference premium/Market premium | 3,558 | 3,383 | 4,678 | 5,514 | 5,396 | 5,253 | 1,095 | 7,720 | 1,016 | 1,459 |

Table 4. 1 shows the premium amounts of each player from the ten-player competitive game. Player/insurer 8 has the highest premium because he has the highest level of risk uncertainty as measured by the claim exposure amount. The player charges a high premium to leverage and minimize risk exposure and potential losses. Player/Insurers 7 and 10 have the least

premium and claim exposure amounts. Other insurers are just setting up competitive premiums to remain in the game. Market premiums are less than the game/NASH premiums except for the one of players 7 and 10 with the most negligible amounts. The bottom section complements the NASH-equilibrium premium table above. It contains the claimed exposure amounts from the game benchmarked against the market. The market is always less than the output from the games. The same approach applies to cooperative games, but in this case, insurers most likely set the same premium as they agree and cooperate and adhere to game rules.

Table 4. 2: Inverse transactions (Insurers-policyholder games) risk versus. Return

| Year | Insurer: Net Written Premium | Policyholder: Net Claims Incurred |
|---|---|---|
| 2008 | 5753 | 3296 |
| 2009 | 5452 | 3110 |
| 2010 | 5649 | 3541 |
| 2011 | 5846 | 3005 |
| 2012 | 5845 | 3177 |
| 2013 | 5772 | 2805 |
| 2014 | 5427 | 2735 |
| 2015 | 5139 | 2779 |
| 2016 | 4397 | 2169 |
| 2017 | 4637 | 2305 |
| 2018 | 4573 | 2730 |
| 2019 | 4173 | 3623 |
| 2020 | 4265 | 3412 |
| 2021 | 4713 | 3079 |

Considering two players from the predator-prey model, we explore the nexus between risk (claims) and return (premiums). We compared the data for the entire period covering 2008 to 2021. We find a negative relationship between the two, indicating competitive game inclusion. Based on the utility theory, both players aim to maximize their utility (self-interest). Policyholders avoid paying high premiums, and insurers are keen to enjoy economies of trading from these high premiums, hence the competition. Using Table 4. 2, we used linear slope regression to quantify the relationship between the net written premiums and net claims incurred. We follow the work on physical quantity association assessment by Animasaun, I. L et al. (2022) and Animasaun, I. L et al. (2020). The results suggest that over time, the insurer (Net written premium) decreases at -130.16 while the policyholder (net claims incurred) decreases at -15.18, mainly due to competing interests. We further elaborate on this idea in Figure 1 below.

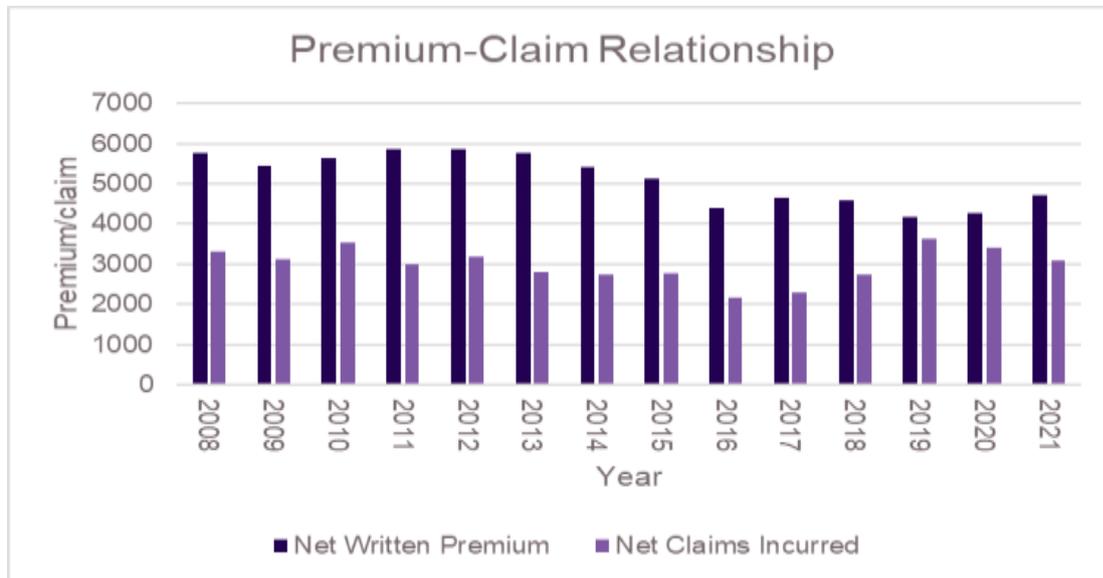

Figure 4. 1: Premium-Claim Relationship

The graph displays the link between written premiums and claims between two games from 2008 to 2021. It shows the return and risk attribution of the two players over time in a well-defined competitive market environment. The stochastic nature of the market explains the market differentials and continued trades/games. Generally, insurers(predator) charges high premiums as they are playmakers, and after trying to avoid these high rates, policyholders end up adhering to the rates.

5. Discussions

This paper derived and explained the competitive and cooperative games that exist in the stochastic non-life insurance markets of South Africa. We provided a novel mathematical approach to deriving the NASH-equilibrium points based on the Lotka-Volterra differential equations. We also derived the optimal trading points from the models for any number of players under competitive and cooperative game environments. Firstly, we considered an insurer-insurer game consisting only of insurers competing or cooperating with premium settings. Each insurer aims to maximize returns by setting high premiums and minimizing claim risk exposures. We found that insurers with high claim exposures tend to set high premiums. The other competitors either set a competitive premium or adopt the fixed premium charge to remain in the game; otherwise, they will operate below the optimal point. We also noted an inverse link between trading premiums and claims in general insurance games due to self-interest and utility indifferences. Fundamentally, since insurers are playing/price makers, policyholders collude, especially when the games among insurers are cooperative. This is because policyholders would have limited alternatives, increasing opportunity costs and minimizing payoffs.

## 6. Conclusion

In this paper, we provided a novel way of deriving and analyzing the NASH equilibrium points to explore and examine the behavior and variation of utility between the two conflicting players in a closed Nash equilibrium loop. We derived a stochastic Lotka-Volterra model from which we fundamentally observed that while an insurer aims to charge high premiums to enjoy more, policyholders are willing to avoid these charges by paying less. Additionally, we analyzed the strength of each player's strategy through stability analysis of each of the derived NASH equivalent points. We noted that a player's strategy is considered superior to the opponent's if it is stable. The trajectories converge towards that critical point, and the opposite holds for weak strategies. We conclude that non-life insurance games are stochastic, and thus, both players need efficient and uncommon strategies to maximize their game payoffs. In the future, we are interested in looking at more randomized games with a reserve component using the Lotka-Volterra model. We shall also derive updated NASH equilibrium points, and empirical analysis shall be done in a more dynamic setup.

8. **Appendices.**

**Zero-interaction case-graphical results.**

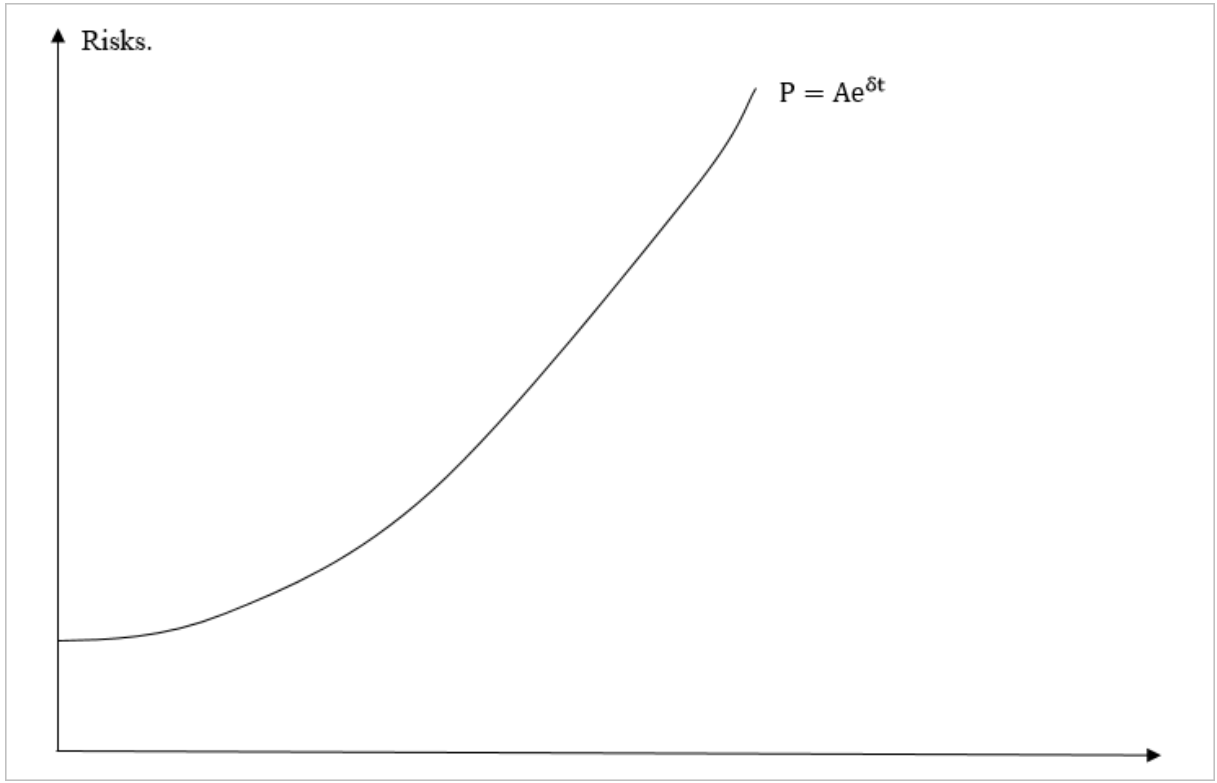

**a). Zero-insurers count case**

The figure shows that, with time, in the absence of insurers, no games will be played, and thus, the risks for policyholders will logistically grow, which is economically unhealthy.

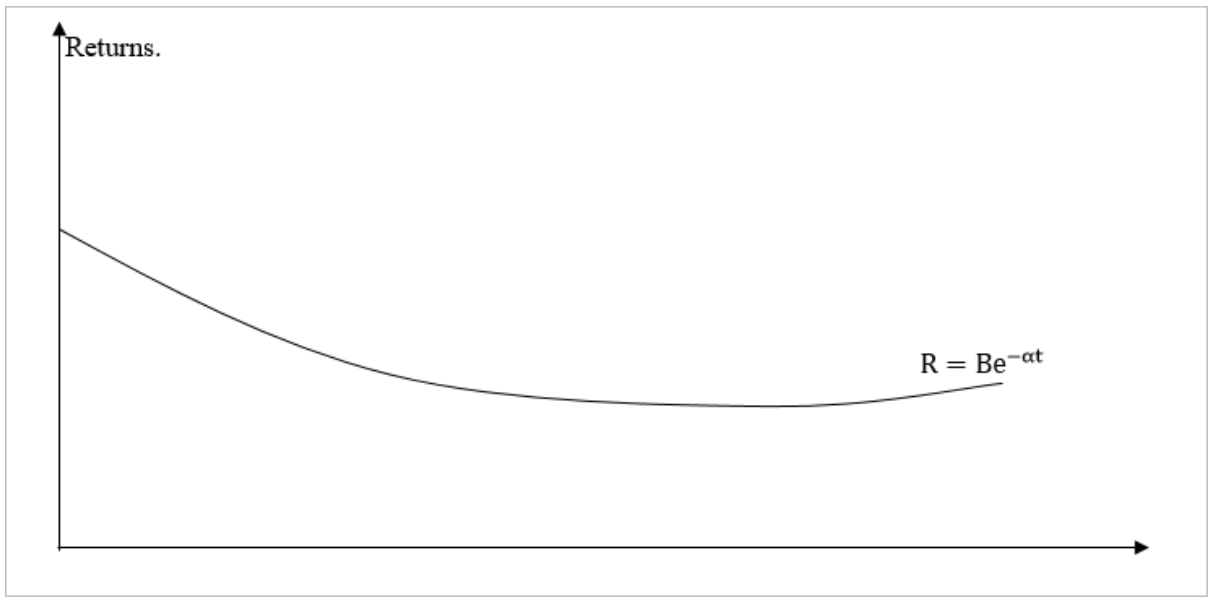

**b). Zero policyholders count case**

The above graph is an accurate inverse representation of the first case. These two functions show the rationality principle within insurance trading games. These functions are for predator-prey and prey-predator models but are applied in inverse considerations.